\documentclass[conference]{IEEEtran}
\IEEEoverridecommandlockouts
\usepackage{cite}
\usepackage{amsmath,amssymb,amsfonts}
\usepackage{algorithmic}
\usepackage{graphicx}
\usepackage{textcomp}
\usepackage{xcolor}

\usepackage{amsmath,amssymb}
\usepackage{algorithmic}
\usepackage{subcaption,siunitx,booktabs}
\usepackage{adjustbox}
\usepackage{graphicx}
\usepackage{tikz}
\usepackage{lscape}
\usepackage{longtable}
\usepackage{blindtext}
\usepackage{tabularx}
\usepackage{lineno,hyperref}
\usepackage{lipsum}
\usepackage[ruled,vlined,linesnumbered]{algorithm2e}

\def\BibTeX{{\rm B\kern-.05em{\sc i\kern-.025em b}\kern-.08em
    T\kern-.1667em\lower.7ex\hbox{E}\kern-.125emX}}
\begin{document}

\title{Solving the optimal stopping problem with reinforcement learning: an application in financial option exercise\\
\thanks{CAPES}
}

\author{\IEEEauthorblockN{1\textsuperscript{st} Leonardo Kanashiro Felizardo}
\IEEEauthorblockA{\textit{Electronic Systems} \\
\textit{Escola Polit\'ecnica, Universidade de S\~ao Paulo}\\
S\~ao Paulo, Brazil  \\
0000-0002-2871-860X}
\and
\IEEEauthorblockN{2\textsuperscript{nd} Elia Matsumoto}
\IEEEauthorblockA{\textit{Sao Paulo School of Economics} \\
\textit{Fundacao Getulio Vargas}\\
S\~ao Paulo, Brazil  \\
0000-0003-1667-4221}
\and
\IEEEauthorblockN{3\textsuperscript{rd} Emilio Del-Moral-Hernandez}
\IEEEauthorblockA{\textit{Electronic Systems} \\
\textit{Escola Polit\'ecnica, Universidade de S\~ao Paulo}\\
S\~ao Paulo, Brazil \\
0000-0003-4554-168X}
}

\maketitle

\begin{abstract}
The optimal stopping problem is a category of decision problems with a specific constrained configuration. 
It is relevant to various real-world applications such as finance and management.
To solve the optimal stopping problem, state-of-the-art algorithms in dynamic programming, such as the least-squares Monte Carlo (LSMC), are employed.
This type of algorithm relies on path simulations using only the last price of the underlying asset as a state representation.
Also, the LSMC was thinking for option valuation where risk-neutral probabilities can be employed to account for uncertainty.
However, the general optimal stopping problem goals may not fit the requirements of the LSMC showing auto-correlated prices.
We employ a data-driven method that uses Monte Carlo simulation to train and test artificial neural networks (ANN) to solve the optimal stopping problem.
Using ANN to solve decision problems is not entirely new.
We propose a different architecture that uses convolutional neural networks (CNN) to deal with the dimensionality problem that arises when we transform the whole history of prices into a Markovian state.
We present experiments that indicate that our proposed architecture improves results over the previous implementations under specific simulated time series function sets.
Lastly, we employ our proposed method to compare the optimal exercise of the financial options problem with the LSMC algorithm.
Our experiments show that our method can capture more accurate exercise opportunities when compared to the LSMC. We have outstandingly higher (above 974\% improvement) expected payoff from these exercise policies under the many Monte Carlo simulations that used the real-world return database on the out-of-sample (test) data.
\end{abstract}


\section{Introduction}
\label{sec:intro}

The optimal stopping problem is relevant topic especially for option exercise (financial and real-options) \cite{Longstaff2001,Hanfeld2017,Nadarajah2017}.
The great variety of underlying assets and the distinct rules of each option created opportunities to explore many computational and mathematical tools for support. 
For instance, one crucial option rule is an early exercise option, such as the American options style. 
This class of options introduces the optimal stopping problem in the option exercise context and opens the doors for many mathematical and computational techniques to solve it \cite{Peskir2006,Egloff2005,Becker2019,Ruf2019}. 
The goal is always to find a policy that can make a stop decision.
Many decision problems can be framed as optimal stopping problems.
The most famous application is the optimal exercise of an option to maximize an arbitrated utility function. 
The stopping problem can be interpreted as an exit or entry decision in an investment such as a buy/sell action or an exercise action. 
In this context, it is common to deal with market uncertainties using real options, giving flexibility for investment and divestment decisions.
Other applications can be many folds, such as financial market \cite{Broadie2000,Longstaff2001,Becker2019}, inventory/production management \cite{Miller2002,Huang2012}, investment decisions on technology/products \cite{Pennings1997}, reservoir management for hydropower \cite{Steinschneider2012}, real state \cite{Bulan2009}, mergers and acquisitions \cite{Hackbarth2008}, venture capital \cite{Li2008}, advertising \cite{Kamrad2005} and environmental compliance \cite{Cortazar1998} (see \cite{Lander1998} for more details).

%When applied for financial option valuation, it is a common assumption that the \textit{risk-neutral} probabilities, which assumes arbitrage-free pricing with linear pricing consistency between all trader securities regarding their payoff components.
%If we use real-world probabilities to price an asset, the expected values of each security would require adjustments for each risk profile.
%It is important to remember that arbitrage-free pricing is not fundamentally the \textit{correct} price; it is only the \textit{market consistent} price.
%Furthermore, the risk-neutral valuation approaches implicitly incorporate the risk premia in the underlying asset's market price.

%Describle approaches that use path simulation such as LSMC
Optimal stopping problems in finance usually assume a time series behavior, such as the geometric Brownian motion correctly describing the underlying asset dynamics.
Methods such as the least-squares Monte Carlo (LSMC) \cite{Longstaff2001} uses path simulations to find optimal stopping points.
According to \cite{Longstaff2001}, the LSMC aims to provide a pathwise approximation to an optimal stopping rule to maximize the value of the American Option.
Many other articles formerly also used the simulation approach \cite{Bossaerts1989,Tilley1993,Carriere1996,Broadie1997} for American options pathwise simulation, employing the stratification and parameterization techniques to approximate the transition state density function or the exercise boundary.
Although the conditional expectations based on the approximated policies are a nonparametric regression, the simulated path used for the regression is not.
The predominant model is the geometric Brownian motions that assume that the returns follow a normal distribution to generate the simulated paths. 
Each price is a Markov state, containing all the information for the subsequent returns.
Assuming a Markovian framework is a practical modeling decision, it does not necessarily reflect the reality of many other problems.
Studies such as \cite{Becker2019,Goudenege2020} provide examples of how machine learning can help to solve the optimal decision problem under non-Markovian structures. 

The main advantage is the non-dependency of multiple simulations and adjustments of the path generation model over the real data.
An advantage of the ANN is the universal functional approximation capability \cite{Kurkova1992}.
Despite the mathematical proofs of the approximation capabilities \cite{Cybenko1989}, the use of better architectures can lead to improvement in specific task performances.
The real-world probabilities may be estimated in some cases, as we further see, leading to better estimation of optimal stopping points.
Finally, ANN can provide a solution for great dimensionality problems, which are sometimes present when converting a non-Markovian state into a Markovian one.
The use of more optimized architectures as the convolutional neural networks (CNN) \cite{Hubel1962,Fukushima1980} brings feature extraction capabilities that help overcome the dimensionality problem.
By using a set of adaptive parameters (filter), the CNN specializes in pattern recognition by learning the kernel coefficients using the back-propagation method \cite{LeCun1998}.
This pattern recognition, which results from the adaptive filters, can synthesize the state with the relevant information for an ANN to approximation exercise policies.

In this context, this paper proposes using convolutional neural networks to solve the optimal exercise problem in the option domain by using the learning procedures described in \cite{Becker2019}. In their work, Becker et al. describe a method to train a neural network to predict an optimal stopping time sequence recursively produced by running a stochastic gradient ascent procedure. 

%State the contributions of the paper (maybe using topics):

We summarize the contributions of this paper as follows:

\begin{itemize}
    \item Improvements on the original implementation of \cite{Becker2019} \textbf{addressing the dimensionality problem} for a broader category of time series functions. 
    \item \textbf{Proposing a new approach} to optimally exercise options in the context of financial options
    \item \textbf{Increasing the optimal stopping performance} in the real-world data by transforming the historical information in a Markov state with the feature extraction of the CNN layer.
    \item We show that, by using a universal function approximation approach such as ANN, we can \textbf{avoid the calibration process} that can carry functional biases for optimal stop policies.
\end{itemize}

In Section \ref{sec:problem definition}, we define our problem, and propose the solution for the problem in Section \ref{sec:solv_probl}. We conduct our experiments in Section \ref{sec:experiments} and show the results. Finally, in Section \ref{sec:conclusion}, we present the conclusions and considerations for future works.

\section{Problem definition}
\label{sec:problem definition}

To hedge risks related to asset price fluctuation financial, institutions create different types of derivatives, which is a contract with an underlying asset. That contract can have many formats, and one of them provides the investor the option to buy or sell the underlying asset for a predetermined price. This type of contract is called an option. The optimal stopping problem is when the investor has to decide the best moment to execute the option to maximize the payoff. The option can be the exercise on any data for a positive payoff, or non-exercised, giving the investor zero value payoff. The payoff of an option depends on the type, put (right to sell in the future) or call (right to buy in the future). Furthermore, there are different options, such as European-style options, whereby the investor cannot early exercise the option, only at maturity. There is the case of the American-style option, whereby the investor can early exercise at any time. For the American (index $a$) and European (index $e$) options, to calculate the payoff, we compare the strike price ($K$), previously agreed price in the contract, and the asset price ($S$). Equation \eqref{eq:european_call}

\begin{equation}
    C_{T}^{e}=\max \left\{S_{T}-K, 0\right\},  \label{eq:european_call}
\end{equation}
expresses the price of the European ($e$ index) call option price, which is the maximum payoff in an period $T$ and analogous to the call option 

\begin{equation}
    P_{T}^{e}=\max \left\{K-S_{T}, 0\right\},  \label{eq:european_put}
\end{equation}
expresses the value of the European put option.

The American-style call option can be early exercised, but this is never optimal for a non-dividend-paying stock. Consequently, in the scenario of non-dividends, the price of a European call option is the same as the American call option. In the case of dividend payments, it may be optimal to exercise the option as it goes ex-dividend (the day that the amount of the dividend is announced). In turn, in the case of the American-style put options, as the value of the put option cannot be less than the intrinsic value, 
\begin{equation}
    P_{T}^{a} \geq K-S_{T},  \label{eq:american}
\end{equation}
since it can be early exercised, the intrinsic value can be larger than the continuation value (not exercising the option). The American-style put option has the optimal stopping point, unlike the call option. Since the value of a stock it is not negative, for European-style options, we have a upper bound defined by the present value of the strike price $PV(K) = Ke^{-r(T-t)}$ and for the American-style put, an upper bound equals $K$ because American-style option can be exercised at any moment. The difference between the price of the European-style put option and the American-style put option is the option time value.
Consequently, the problem is to find the critical point where the intrinsic value surpasses the continuation value of an American-style put option, or in other terms, the point where the time-value is below zero. Fiding the critical point is a challenging problem since we do not know the stock price movement. To find the optimal stopping point, we require finding the boundary region to exercise the option, which can be solved as an optimization problem. This optimization problem gets more complicated as we increase the complexity of the option mechanism, including more underlying asset and complex rules that govern the put price.

One popular option contract used in the market is the Bermudan style option, in which it is only possible to exercise the option at pre-determined times. 
This type of contract is usually interpreted as a discretization of the American style options. 
As a method of approximating the market decision process, discretization of time is one common approach.
Here try to solve the optimal stopping problem for the Bermudan style options using universal function approximation methods such as artificial neural networks.
By analyzing state variable $s$, composed of the underlying asset price return,  interest $r$, strike price $K$, and payoff $p$, the model returns a decision (action) $a$ at each decision step.
The optimal stopping problem of the form $\sup _{t} \mathbb{E} g\left(t, s_{t}\right)$ where $s=\left(s_{n}\right)_{n=0}^{N}$ where $N$ is the number of time-steps, $s\in S$ is an state, and $t\in T$ is the stopping time in the set of all stopping times. 
The definition given covers the case of Bermudan options in which we have finite stopping decisions. However, this definition is also relevant for the continuous case since its time-discretized version can approximate it.

The Brownian motion is one common assumption for the movement of asset prices in financial markets. For many reasons, it is convenient, i.e., option pricing.
In this work, we are concerned only with the stopping problem from the perspective of the option owner.
It tries to approximate the real probabilities function from the probability space $(\Omega, \mathcal{F}, P)$, and horizon $[0,T]$, with $\Omega$ the sample space, $\mathcal{F}$ the sigma field of distinguishable events at time $T$, $P$ the probability function for each event in the event space.
When solving the stopping problem, if the increments are correlated by following an underlying function, we need to keep track of all past movements to have a Markov state.
The dimension of the problem is equal to the number of time-steps $N$ used to approximate the continuos-time.
For that reason, as we try to solve the problem for longer time horizons with more precise time-discretization, we may suffer from a dimensionality problem.

We investigate our proposed technique for the optimal exercise problem of the put options model as an optimal stopping problem case. 
It is possible to employ an American-style options model to solve various problems such as energy contracts. 
One example is the use option to contract the delivery of  $q$ crude oil at a price $p$, resulting in an investment of $K$ (or strike price), at time $T$. 
This contract has two parts; the purchase part can buy the total value in crude oil $I$ at each time $t\in T$ buy the price $K$ price. The selling part endows the buyer with $n\leq N$ exercise rights. 
The problem presented can be modeled as a Bermudan option and solved using techniques such as the least squared Monte Carlo (LSMC).
That type of contract in energy markets is just an example of a more extensive range of contracts in the industry under the label of \textit{option to contract}.
An over risk-free premium is required for real options and can be acquired by optimally exercising the option.

\section{Solving the optimal stopping problem}
\label{sec:solv_probl}

To solve the long time horizon of Bermudan style options, we propose a reinforcement learning approach using an artificial neural network architecture that can capture the long-term time-dependent functions. 
Also, we try and overcome the dimensionality problem when we try to bring to the Markov framework by using the whole historical data.
For this, we develop an decision algorithm based on the method proposed by \cite{Becker2019}, changing the policy learner neural network by convolutional neural networks and using the history of price returns (not only prices as employed by \cite{Becker2019}).
\cite{Becker2019} approach is direct policy ($\pi$) optimization the equation \eqref{eq:loss_net} as a loss function,

\begin{equation}
\sum_{n=1}^N p \pi_\theta(s_n) - g_t(1-\pi_\theta(s_n)),
\label{eq:loss_net}
\end{equation}
where $p$ is the option's payoff, $\theta$ is the parameters, and $g$ is the last payoff relative to the stop action. 
The process to approximate policy occurs iteratively starting from the terminal stopping decision $\pi_\theta(s_N)\equiv 1$, being the stopping policy approximated by ANN $\pi_\theta:\mathbb{R}^{d} \rightarrow\{0,1\}$. 
This process proceeds backward in time, from the terminal state to the first state.
The neural network approximate try is to approximate the optimal stopping policy $\pi^*$ by the $\pi_\theta$ which generates the stopping decision $[0,1]$ by the forward computation where we get \textbf{the output $v$ of the ANN} by

\begin{equation}
    \ v_{j}^{(l)}(s)=\sum_{i} w_{j}^{(L)}(s) y_{i}^{(l-1)}(s)
\end{equation}
where $l$ is the neural network layer, $y_i$ is the output of neuron $i$ previous layer, $w$ is synaptic weights that compose parameters $\theta$ with bias $b$. $y_i$ is the result of the last output

\begin{equation}
    \ y_{j}^{(l)}=\varphi_{j}\left(v_{j}(n)\right)
\end{equation}
where $\varphi$ is the activation function, such as the ReLU or sigmoid.
The weights of the network are updated using the loss described in Equation \ref{eq:loss_net}.

\begin{equation}
    \ w_{j i}^{(l)}(n+1)=w_{j i}^{(l)}(n)+\alpha\left[\Delta w_{j i}^{(l)}(n-1)\right]+\eta \delta_{j}^{(l)}(n) y_{i}^{(l-1)}(n)
\end{equation}
where $delta$ is the local gradient, $\eta$ is the learning rate and $\alpha$ is the momentum (check \cite{Haykin2009} Chapter 4 for more details).
Under the assumption of a geometric Brownian motion simulation path, we would not require using the whole history of the price time series to generate the path simulation.
n policy $\pi$ has parameters $\theta$
When we enter a series that can be auto-correlated, we require the whole history of prices to estimate the optimal stopping point.
Fortunately, \cite{Becker2019} brief explores this possibility using the fraction Brownian motion to generate the time series and adapts his implementation to receive the whole price time series as input.

\begin{figure*}[ht]
    \centering
    \includegraphics[width=0.95\linewidth]{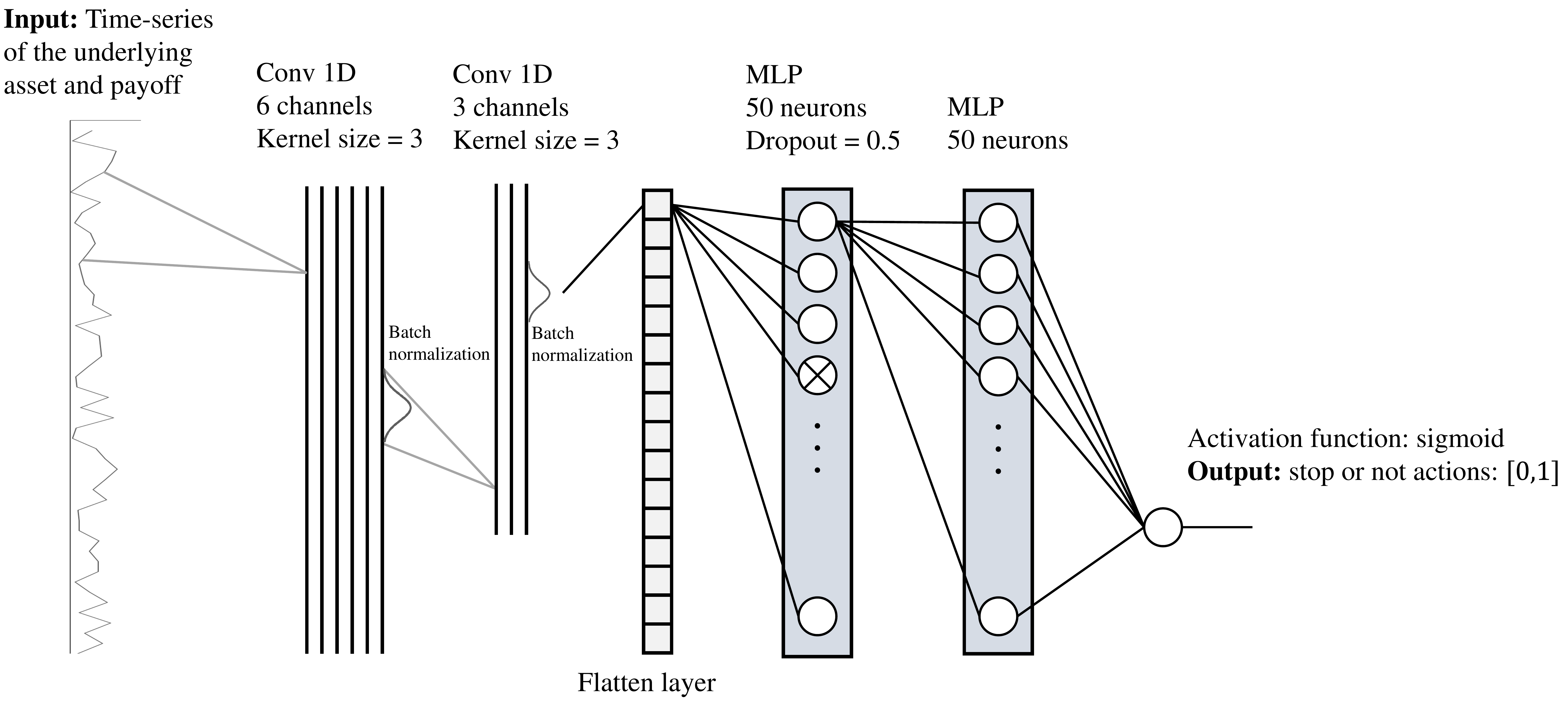}
    \caption{Network architecture}
    \label{fig:network}
\end{figure*}

The main problem is that the multilayer perceptron architecture (MLP) has limitations related to overfitting the noisy signal.
Up to this point, we focused on the multilayer perceptrons (MLP) formulation, but we employed a particular class of ANN called convolutional neural networks (CNN) \cite{Hubel1962}.
Initial studies conceived the CNN to recognize two-dimensional shapes with invariance to translation, skewing, scaling, and other distinct distortions.
As it is done for the MLP, the CNN learn in an supervised manner following the constrains on the \textit{feature extraction}, \textit{feature mapping} and \textit{subsampling} \cite{LeCun1998}.
We depict the proposed architecture for the optimal exercise problem in Fig. \ref{fig:network}.
First, we have textbf{input the set of time series considered to estimate the optimal stopping point} where the first hidden layer performs convolution.
The first hidden layer consists of six feature maps, with each feature map comprising of 23 neurons, each neuron with a receptive field of size 3.
The second hidden layer is similar to the first but with 22 neurons and a receptive size of 3.
The third layer performs a reshape operation to serve as input for the subsequent MLP.
The fourth and the fifth layers are dense layers, also called fully connected layers, with 50 neurons each.
Our proposed architecture makes it possible to include multiple time series features as input, including each time series in a different channel, dealing with multiple time series as inputs.

\IncMargin{1em}
\begin{algorithm}
\SetKwData{Left}{left}\SetKwData{This}{this}\SetKwData{Up}{up}
\SetKwFunction{Union}{Union}\SetKwFunction{FindCompress}{FindCompress}
\SetKwInOut{Input}{input}\SetKwInOut{Output}{output}

\Input{$S_0$,$d$,$K$, $N$, $r$, $\sigma$, $\delta$}
\Output{Trained parameters: $w$}

initialization: model NN parameters: $w_0$, max number of episodes: $max\_ep$ ,batch\_size: $b$, choose the type of data generation: $gen$, $max_price$ = $0$

\ForEach {episode in range(0,$max\_ep$)}{
$X,p,r,K\longleftarrow gen(b,S_0,d,K,N,r,\sigma,\delta)$\;
$g_{t} \longleftarrow p[:,-1]$\;
$s = concatenate(X,p,r,K)$\;
$hist_a = []$\;
\ForEach {step in range($N-2$,$1$,$-1$)}{
$a \longleftarrow NN(w,s[step])$\;
$hist_a.append(a)$\;
$loss -= mean_b(p \times a - g_{t}\times (1-a))$\;
$g_{t} = where(a>0,p,g_{t})$\;
}
$opt\_price \longleftarrow mean(g_t)$\;
$w \longleftarrow BackPropagation(w,loss)$\;
}

\caption{Training the CNN for the optimal stopping decision problem}\label{algo:LSTM_train}
\end{algorithm}\DecMargin{1em}

\subsection{The baseline method: LSMC}
\label{subsec:baseline}

To compare our proposed method in an real-world case, we use a the LSCM, a standard baseline method to solve the optimal stopping problem.
The LSMC is a mathematical approach based on simulation to approximate the American style options value.
In the LSMC method we regress the ex post realized payoffs from continuation on functions of the values of the state variables.
With the conditional expectation function we can obtain the optimal exercise strategy along each path \cite{Longstaff2001}.

We can list a few reasons why the LSMC may be not appropriate for the real-option optimal stopping problem:

\begin{itemize}
    \item The LSMC tries to approximate the conditional expectation using a linear regression which can be inadequate since we are trying to approximate an arbitrary function that can be highly non-linear.
    \item In a non-Markovian model, we may have to project a high dimensional state variable into a vector of parameters $\beta$ with hundreds of entries. 
    \item Monte Carlo methods are model-dependent (model-based), which means we need to know the Markov decision process (MDP) to optimize the action. Sometimes we may not know the MDP, or it may be too complex to model easily.
\end{itemize}

Despite these arguments, the LSCM is well known for its broad applicability; hence, it can be considered a reasonable benchmark.

\section{Experimental verification}
\label{sec:experiments}
We divide the experiments into two phases: First, we compare our approach against the \cite{Becker2019}'s approach for different time series generation functions assuming that we are dealing with the case in which the asset price return does not depend only on the last asset price; In the second batch of the test, we use our proposed approach to solve the optimal exercises/stopping problem for real-option with the underlying asset being an energy market asset. To reproduce the results, we provide the code in:  \url{to_be_included_after_double_blind_review}. We implemented all the algorithms using the \textit{Python} language with \textit{Pytorch} and executed it in a PC containing an \textit{AMD Ryzen 5 5600X 6-Core} processor 3.70 GHz, an \textit{NVIDIA} \textit{GeForce} \textit{RTX 3060 GPU}, and \textit{32GB RAM}.

\subsection{Analysis for the path simulation approach}
\label{subsec:simults}

\begin{figure*}[ht]
\begin{subfigure}{.5\textwidth}
    \centering
    \includegraphics[width=1\linewidth]{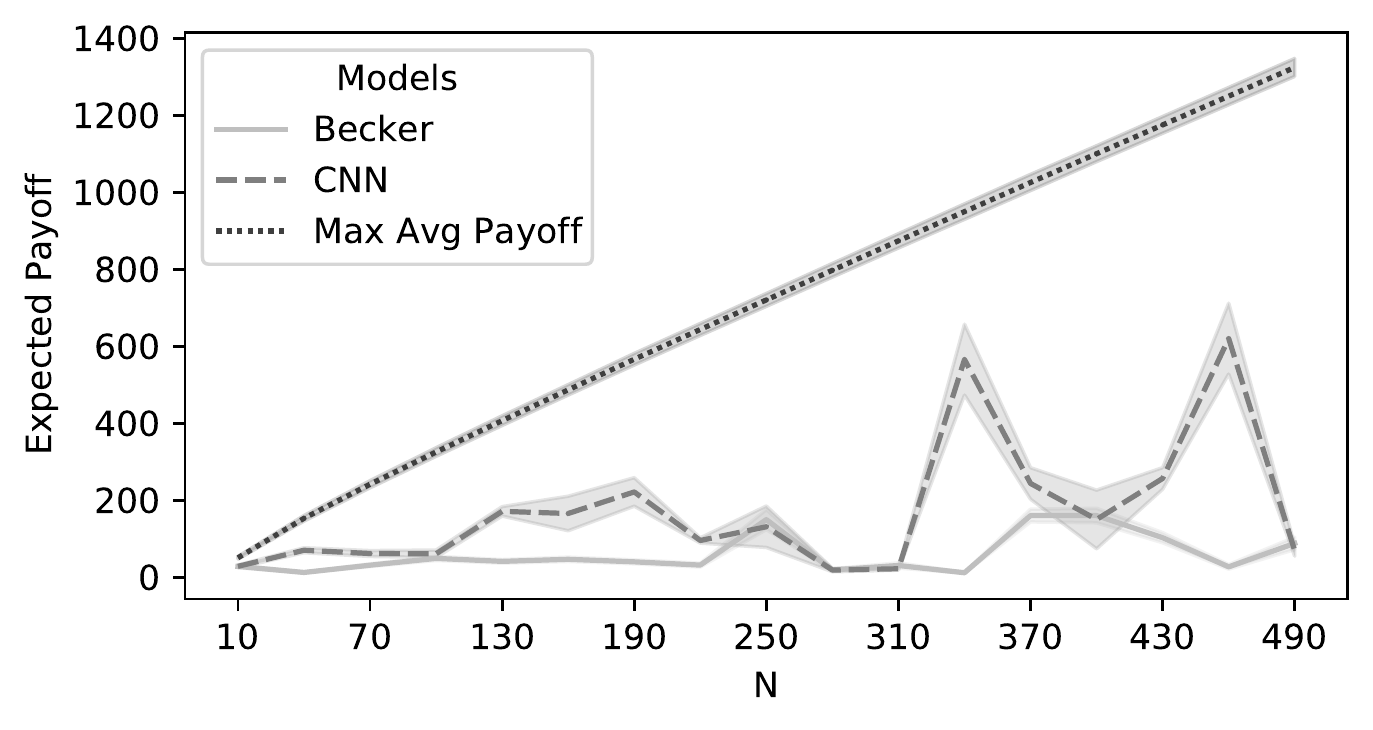}
    \caption{Average payoff obtained with Becker's model and ours for each value of N (maximum number of Bermudan steps). The stock prices are simulated by a harmonic composition of two sines and Gaussian noise. We represent the standard deviation with the filled areas.}
    \label{fig:harmonic_results}
\end{subfigure}%
\begin{subfigure}{.5\textwidth}
    \centering
    \includegraphics[width=1\linewidth]{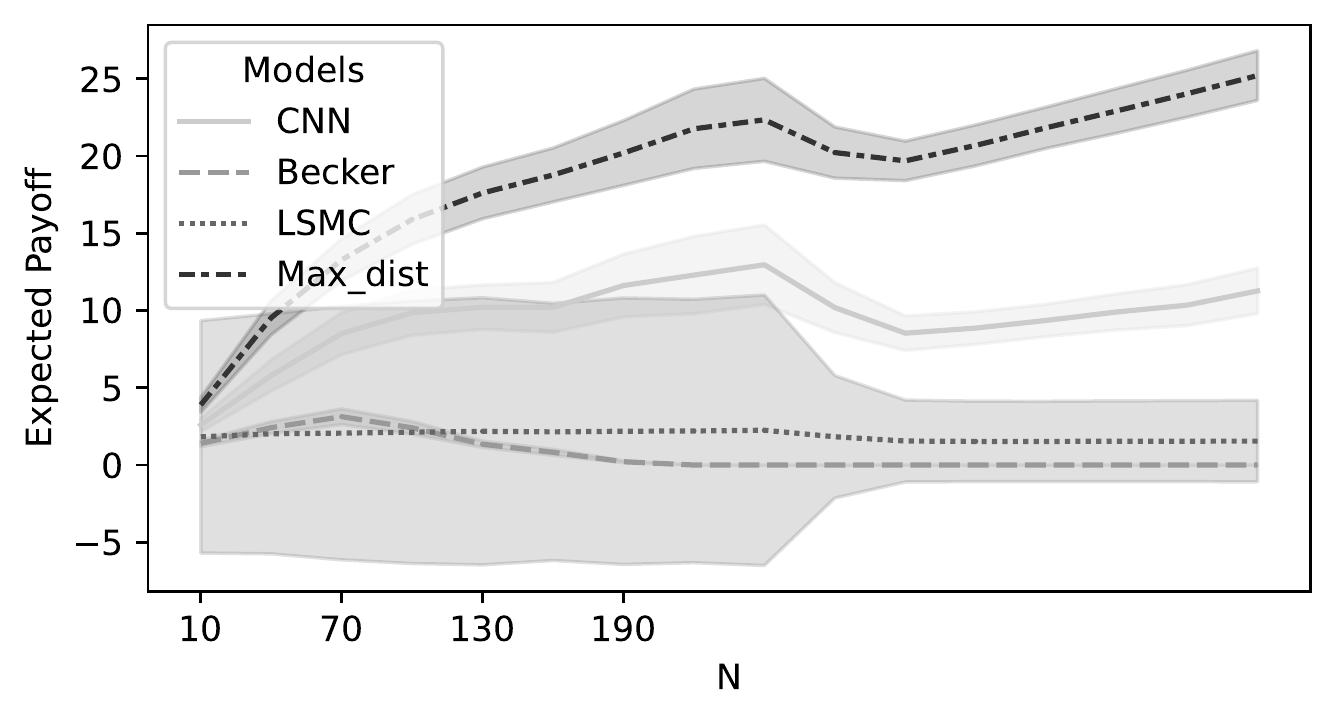}
    \caption{Average payoff obtained with Becker's model and ours for each value of N (maximum number of Bermudan steps). We present the simulation using samples of returns obtained using oil price.Also, we represent the standard deviation with the filled areas.}
    \label{fig:energy_results}
\end{subfigure}
\begin{subfigure}{.5\textwidth}
    \centering
    \includegraphics[width=1\linewidth]{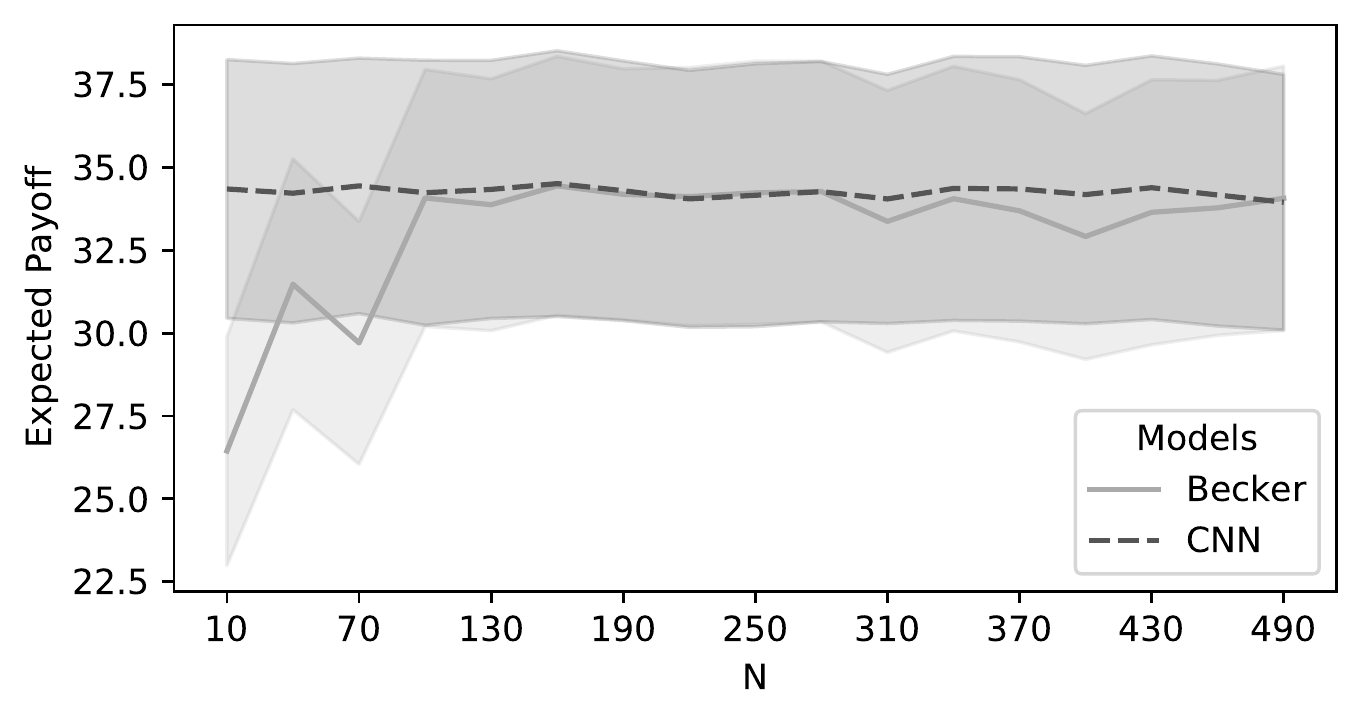}
    \caption{Average payoff obtained with Becker's model and ours for each value of N (maximum number of  Bermudan steps). The stock prices are simulated by fractional Brownian motion. Here the filled area represents the 95\% confidence interval for better visualization.}
    \label{fig:fbm_results}
\end{subfigure}
\begin{subfigure}{.5\textwidth}
    \centering
    \includegraphics[width=1\linewidth]{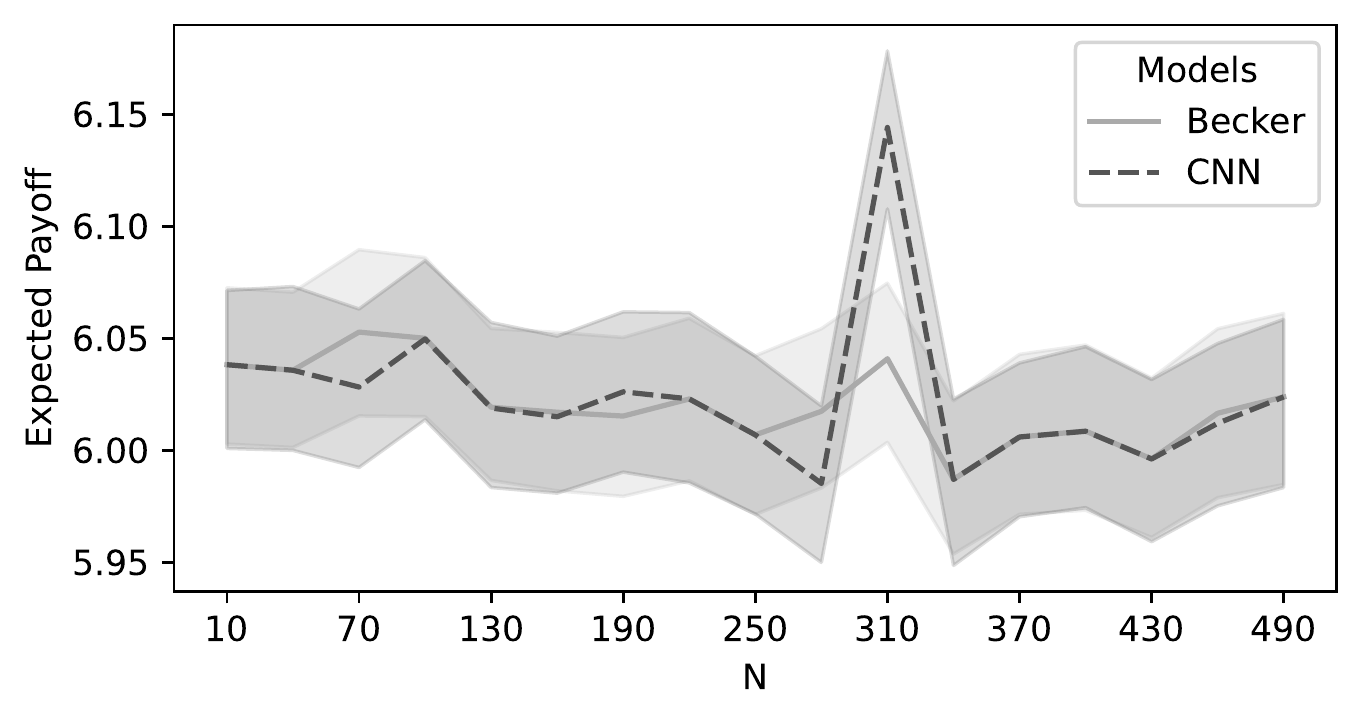}
    \caption{Average payoff obtained with Becker's model and ours for each N value (maximum number of Bermudan steps). The geometric Brownian motion simulates the stock prices. Here the filled area represents the 95\% confidence interval for better visualization}
    \label{fig:brownian_results}
\end{subfigure}
\caption{CNN and Becker's model comparison for different time series}
\label{fig:all_figures}
\end{figure*}

As the first batch of experiments, we compare our CNN approach to \cite{Becker2019}'s approach using three different path generations: geometric Brownian motion, fractional Brownian motion, and harmonic composition with Gaussian noise. We calculate the average payoff in $326000$ Monte Carlo simulations (paths). We specify the simulation parameters for each generation technique in Table \ref{tab:sythetic},

\begin{table}[ht]
\centering
{\small
\caption{Simulation generator parameters}
\label{tab:sythetic}
\begin{tabular}{ll}
\toprule
\textbf{Generator}         & \textbf{Parameters}\\
\toprule
\begin{tabular}[c]{@{}l@{}}Geometric \\ Brownian motion\end{tabular}  & \begin{tabular}[c]{@{}l@{}}$S_0 = 120; K = 100; T = 3; r = 0.05;$ \\ $\delta = 0.1; \sigma = 0.1$\end{tabular} \\\hline
\begin{tabular}[c]{@{}l@{}}Fractional \\ Brownian motion\end{tabular} & \begin{tabular}[c]{@{}l@{}}$S_0 = 120; K = 100; T = 3; r = 0.05;$ \\ $\sigma = 0.1, h = 0.7$\end{tabular}  \\\hline
\begin{tabular}[c]{@{}l@{}}Harmonic + \\ Gausian noise\end{tabular}   & \begin{tabular}[c]{@{}l@{}}$S_0 = 120; K = 100; T = 3; r = 0.05;$\\  $ampl = 0.2; freq_1 = 0.3; freq_2 = 2$\end{tabular} \\
\bottomrule
\end{tabular}}
\end{table}
where $S_0$ is the initial asset price, $K$ is the strike price, $T$ is the number of years, $r$ is the risk-free rate, $\delta$ is the dividend rate payment, $\sigma$ the constant volatility value, $h$ is the Hurst parameter used for fractional Brownian motion, $ampl$ is the amplitude of both harmonics, and $freq$ is the frequency, that is different for each harmonic.
We explore how the exercise steps affect the overall results of both methods.Fig. \ref{fig:brownian_results}, Fig. \ref{fig:harmonic_results}, and Fig. \ref{fig:fbm_results} depict how the average payoff evolves comparing our CNN approach against Becker's original implementation. The comparison uses three different time series generation techniques.

Fig. \ref{fig:brownian_results} shows that our method can achieve the same performance as the Becker implementation. 
We also explore the performance of the training phase for both implementations. 
Note that both algorithms stabilize in a few epochs when dealing with the geometric Brownian motion series and also that the exercise occurs mainly on the last exercise opportunity.
The exercise concentrates on the last decision step if we choose a specific set of path simulation parameters.
The Brownian motion parameters and exercise regions can be better explored (as in many other studies and is a branch of research \cite{Villeneuve1999,Lai2004}) to understand why the networks choose some of the early exercise points.
We did not notice any improvement or decrease for the training speed and average payoff by using the CNN approach for the Brownian motion simulation.

\begin{figure}[ht]
    \centering
    \includegraphics[width=1\linewidth]{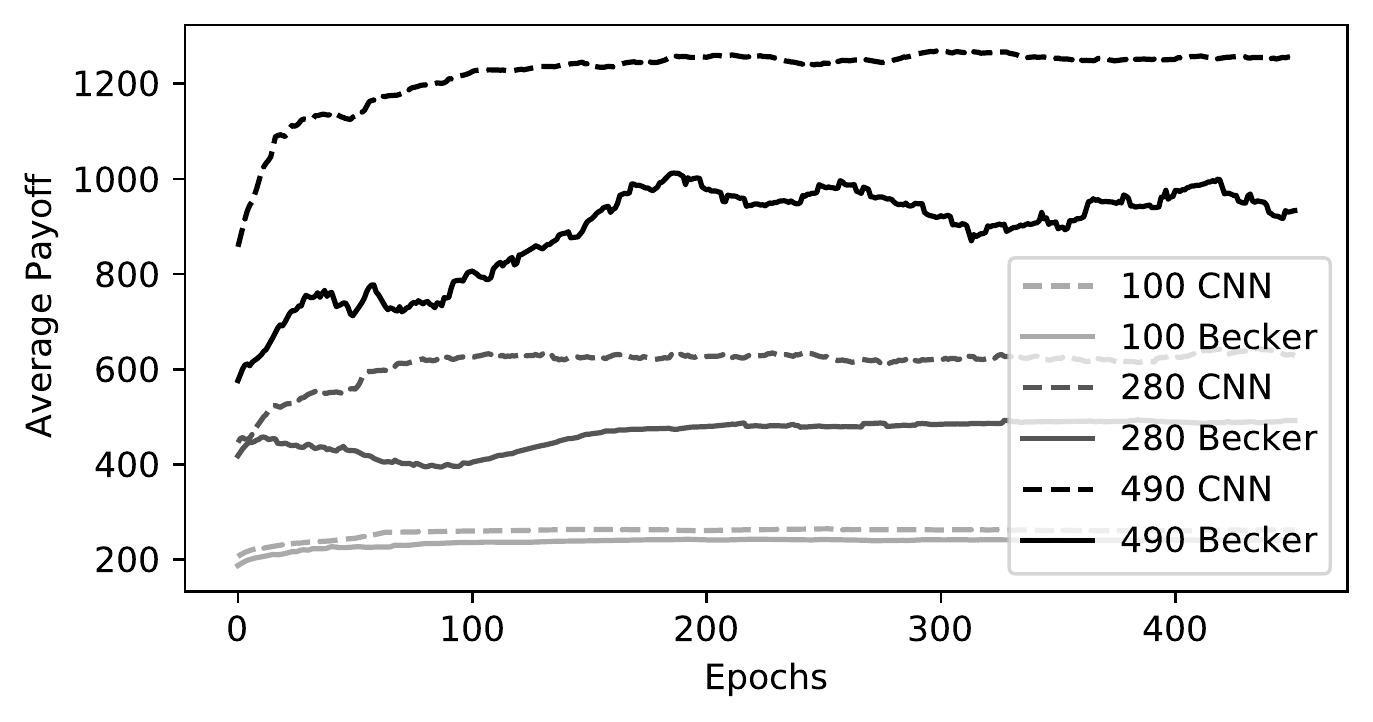}
    \caption{Average payoff versus number of epochs on the training phase.}
    \label{fig:training_harmonic}
\end{figure}

We also employed the fractional Brownian motion in our comparison with the Hurst parameter equal to 0.7. 
In Fig. \ref{fig:fbm_results} we observe that the difference between both approaches (ours and the baseline) is not significant. 
In Fig. \ref{fig:fbm_results}, instead of the standard deviation, we use a 95\% confidence interval, which is better to visualize since the standard deviation is significant and distorts the scale. 
When analyzing the training performance evolution by epoch, we do not perceive any improvement or decrease in learning speed.

In Fig. \ref{fig:harmonic_results}, note that as the dimensionality grows, tne more prominent is the convergence speed difference between our approach and Becker's. 
Furthermore, Fig. \ref{fig:training_harmonic} evidences that CNN implementation can achieve higher performances and much faster than the previous implementation. 
By covering a more extensive group of functions, we recommend using our method to evaluate optimal stopping points for auto-correlated time series.

\subsection{Optimal stopping for real options}
\label{subsec:real options}

\begin{table*}
\centering
\caption{Results with the returns statistics (mean, standard deviation, skewness, and kurtosis) and the respective results of our proposed model and the baseline (LSMC) grouped by sector.}
\label{tab:sector_result}
\begin{tabular}{lllll}
\toprule
       \textbf{Economic Sector} &            \textbf{Mean Return} &\textbf{CNN Expected Payoff} & \textbf{LSMC Expected Payoff} & \textbf{\% Improvement} \\
\midrule
        Communications & -1.31e-03$\pm$0.59 &       4.54$\pm$0.55 &        0.73$\pm$1.37 &    668$\pm$40 \\
Consumer Discretionary &  2.79e-03$\pm$0.61 &       7.59$\pm$1.05 &        0.83$\pm$1.51 &    944$\pm$69 \\
                Energy & -7.90e-04$\pm$1.04 &       6.30$\pm$0.77 &        0.96$\pm$1.73 &    679$\pm$45 \\
             Financial &  2.31e-02$\pm$1.53 &       6.74$\pm$0.66 &        0.60$\pm$1.30 &   1134$\pm$51 \\
             Materials &  9.95e-03$\pm$0.89 &      10.88$\pm$1.33 &        1.38$\pm$2.39 &    833$\pm$57 \\
            Technology &  3.52e-02$\pm$1.01 &      14.35$\pm$1.35 &        0.93$\pm$1.84 &   1586$\pm$73 \\
\bottomrule
\end{tabular}
\end{table*}

In this part of the experiments, we simulate the returns based on the returns present on real-world asset returns data. We obtained the day-frequency price returns for the crude oil asset from NYMEX futures prices. 
We divide the data set into in-sample, with 7658 data points, and out-of-sample (testing) data that we use for evaluation, with 1915 data points. 
In the in-sample data, we have a training and validation data division to control the overfitting of our model. 
We have a 0.7/0.3 proportion between training and validation data for the training set, giving 5630 points for training and 1764 data points for validation.

We employ the same framework used for the simulated data to perform the experiments for real-world data. The main difference is that all the price returns are not generated but randomly picked from real-world data. 
We use these real-world returns to train our network instead of the simulated data. We build each path used for training given by $\{S_0,S_0\rho^p_N,S_0\rho^p_{N-1},...,S_0\rho^p_i\}$, where $i$ is a random number between the training max size minus $N$ and 0, and $\rho^p$ is the relative return.
We defined the required parameters of the experiments as: $S_0 = 100; K = 100; r = 0.05;N=100$.

In the first set of experiments, we test different exercise opportunity sizes and measure the average payoff in the testing data.
We used the LSMC as a baseline model to evaluate our proposed method to solve the optimal stopping.
The LSMC uses the discounted values to decide the optimal stopping point and fit the polynomial, which is not available in the out-of-sample data.
Therefore, there is a gap to fairly compare both our proposed method and the LSMC.
Suppose we use the out-of-sample data to fit the polynomial. In that case, we will be giving a fictional privileged scenario to the LSMC to find the optimal stopping points, even though we test in this scenario.
The hypothesis is that we have underlying hidden functions in the time series, requiring more than just the last price and payoff to estimate the optimal stopping point.
In other terms, that is a non-Markov state that we try to transform into a Markov state by giving history.
The line graph presented in Fig. \ref{fig:energy_results} shows what the maximum theoretical stopping point, our approach, and the LSMC calculated stopping point using out-of-sample data.

We also tested the LSMC using the training data to approximate the polynomials and test these approximated polynomials on the testing data. 
As expected, the results show worse results than for the LSMC as the test performed out-of-sample data directly. 
We used the LSMC trained under in-sample data for the following tests and tested for out-of-sample data.

We tested this optimal stopping problem with put options for any underlying asset to extend the results on financial options.
In this batch of experiments, we try to create a proxy for the possible underlying asset under different scenarios of financial options using stock prices from different economic sectors (communications, consumer discretionary, energy, financial, materials, and technology). 
For each economic sector, we selected five assets and divided them into an in-sample (training and validation) set, with 3779 data points and an out-of-sample (testing) set with 1620 data points.
Here, we try to show that despite the nature of the underlying stock, our method can, in most cases, outperform the LSCM method for optimal stopping decisions.
Table \ref{tab:assets_result} shows the summary of results from our method and the baseline in different assets for distinct sectors.
We can already perceive a significant increase in the average expected payoff from the table by using the CNN approach outperforming the LSMC in all the assets.
The asset with the highest payoffs is also the underlying asset on which both models had the best performances.
Table \ref{tab:sector_result} we can extend the analysis to an economic sector comparison, which maintains the previous analysis.
The Table \ref{tab:sector_result} groups the results from Table \ref{tab:assets_result} by calculating the mean value.
The maximum expected payoff of our approach occurs for the technology sector, which also has the highest mean underlying asset return.
However, LSMC has the highest expected payoff on the material sector, not the sector with the highest asset returns.

\begin{table*}
\centering
\caption{Results with the returns statistics (mean, standard deviation) and the respective results of our proposed model and the baseline (LSMC)}
\label{tab:assets_result}
\begin{tabular}{llllll}
\toprule
\textbf{Economic Sector} & \textbf{Asset} &     \textbf{ Mean Return} & \textbf{CNN Expected Payoff} & \textbf{LSMC Expected Payoff} &\textbf{ \% Improvement} \\
\midrule
        Communications &   BCE &  0.002$\pm$0.50 &       3.07$\pm$0.32 &        0.37$\pm$0.84 &    826$\pm$38 \\
        Communications &    VZ &  0.006$\pm$0.61 &       3.36$\pm$0.36 &        0.79$\pm$1.30 &    424$\pm$28 \\
        Communications &     T & -0.003$\pm$0.47 &       3.55$\pm$0.40 &        0.66$\pm$1.26 &    539$\pm$32 \\
        Communications &   IPG &  0.008$\pm$0.41 &       6.67$\pm$0.81 &        0.64$\pm$1.27 &   1042$\pm$64 \\
        Communications &  DISH & -0.019$\pm$0.94 &       6.07$\pm$0.85 &        1.19$\pm$2.16 &    510$\pm$39 \\
Consumer Discretionary &  ABEV & -0.002$\pm$0.10 &       4.97$\pm$0.55 &        1.15$\pm$1.89 &    433$\pm$29 \\
Consumer Discretionary &   LEN &  0.030$\pm$1.34 &       9.65$\pm$1.20 &        0.74$\pm$1.43 &   1310$\pm$84 \\
Consumer Discretionary &     F & -0.000$\pm$0.20 &       7.18$\pm$1.07 &        0.67$\pm$1.23 &   1078$\pm$87 \\
Consumer Discretionary &   BTI & -0.009$\pm$0.75 &       3.97$\pm$0.48 &        0.58$\pm$1.02 &    687$\pm$47 \\
Consumer Discretionary &   GPS & -0.006$\pm$0.66 &      12.20$\pm$1.96 &        1.01$\pm$1.97 &  1212$\pm$100 \\
                Energy &    BP & -0.005$\pm$0.61 &       4.85$\pm$0.65 &        0.66$\pm$1.32 &    736$\pm$49 \\
                Energy &   COP & -0.002$\pm$1.17 &       7.83$\pm$1.00 &        1.18$\pm$2.15 &    663$\pm$47 \\
                Energy &   CVX &  0.002$\pm$1.80 &       5.07$\pm$0.63 &        0.61$\pm$1.27 &    827$\pm$49 \\
                Energy &   COG & -0.008$\pm$0.50 &       5.05$\pm$0.65 &        1.18$\pm$1.95 &    427$\pm$33 \\
                Energy &   LNG &  0.010$\pm$1.14 &       8.68$\pm$0.92 &        1.17$\pm$1.96 &    741$\pm$47 \\
             Financial &  ACNB &  0.005$\pm$0.70 &       6.12$\pm$0.63 &        0.70$\pm$1.61 &    875$\pm$39 \\
             Financial &   ORI &  0.007$\pm$0.31 &       5.80$\pm$0.59 &        0.53$\pm$1.18 &   1097$\pm$50 \\
             Financial &    RE &  0.047$\pm$3.56 &       5.66$\pm$0.48 &        0.51$\pm$1.05 &   1104$\pm$46 \\
             Financial &   AXP &  0.045$\pm$1.87 &       7.72$\pm$0.69 &        0.56$\pm$1.23 &   1375$\pm$56 \\
             Financial &     C &  0.012$\pm$1.19 &       8.40$\pm$0.91 &        0.69$\pm$1.42 &   1219$\pm$64 \\
             Materials &    DD &  0.010$\pm$1.33 &       6.76$\pm$0.83 &        0.92$\pm$1.67 &    738$\pm$50 \\
             Materials &   BHP &  0.016$\pm$0.95 &       9.37$\pm$0.93 &        1.86$\pm$3.13 &    504$\pm$30 \\
             Materials &  SCCO &  0.020$\pm$0.89 &      11.60$\pm$1.17 &        0.99$\pm$1.72 &   1172$\pm$68 \\
             Materials &   OLN &  0.014$\pm$0.64 &      15.64$\pm$2.23 &        1.42$\pm$2.51 &   1103$\pm$89 \\
             Materials &   MOS & -0.009$\pm$0.66 &      11.03$\pm$1.46 &        1.70$\pm$2.94 &    648$\pm$50 \\
            Technology &  INTC &  0.013$\pm$1.01 &       7.38$\pm$0.73 &        0.63$\pm$1.32 &   1168$\pm$55 \\
            Technology &   HPQ &  0.007$\pm$0.40 &       9.61$\pm$0.99 &        0.67$\pm$1.47 &   1428$\pm$67 \\
            Technology &   AMD &  0.051$\pm$1.19 &      31.41$\pm$3.01 &        2.16$\pm$4.01 &   1452$\pm$75 \\
            Technology &  LOGI &  0.070$\pm$1.16 &      15.69$\pm$1.25 &        0.60$\pm$1.22 &  2599$\pm$102 \\
            Technology &   ARW &  0.037$\pm$1.26 &       7.68$\pm$0.80 &        0.60$\pm$1.19 &   1281$\pm$67 \\
\bottomrule
\end{tabular}
\end{table*}

\section{Conclusions}
\label{sec:conclusion}

In this article, we proposed a direct policy approximation based on \cite{Becker2019}'s architecture but using a better ANN to overcome problems related to high dimensional states spaces. We compared and improved the original baseline model using synthetically generated time series using customized and improved ANN techniques. We applied this solution to a real-world set of problems showing the potential to better exercise real and financial options achieving outstanding results (974\% increase) compared with the literature performance baseline. This result can be extended to any problem related to optimal stopping in the financial and non-financial literature. The main gain of our proposed model may be in the realm of real options where the underlying asset has a behavior that can be predicted but is sometimes non-linear following a wide range of possible functions.
As an extension of this work, we propose using this same architecture in the optimal execution problem using not only the historical prices but the whole history book of negotiations. Our approach seems to be a powerful way to approximate policy function in the finite horizon problem of optimal stopping. Therefore, future researchers can use this algorithm to solve a broader set of problems.

\section*{Acknowledgment}

To be included after double-blind revision

To be included after double-blind revision

To be included after double-blind revision

To be included after double-blind revision

To be included after double-blind revision

To be included after double-blind revision

\bibliographystyle{IEEEtran}
\bibliography{IEEEfull,refs}

\end{document}